\overfullrule=0pt
\input harvmac
\def\a{{\alpha}}
\def\ad{{\dot a}}
\def\bd{{\dot b}}
\def\l{{\lambda}}
\def\b{{\beta}}
\def\g{{\gamma}}
\def\d{{\delta}}
\def\e{{\epsilon}}
\def\s{{\sigma}}

\def\half{{1\over 2}}
\def\p{{\partial}}

\def\t{{\theta}}

\Title{\vbox{\hbox{IFT-P.000/2000 }}}
{\vbox{
\centerline{\bf Cohomology in the Pure Spinor
Formalism for the Superstring}}}
\bigskip\centerline{Nathan Berkovits\foot{e-mail: nberkovi@ift.unesp.br}}
\bigskip
\centerline{\it Instituto de F\'\i sica Te\'orica, Universidade Estadual
Paulista}
\centerline{\it Rua Pamplona 145, 01405-900, S\~ao Paulo, SP, Brasil}

\vskip .3in
A manifestly super-Poincar\'e covariant formalism for the superstring
has recently been constructed using a pure spinor variable. Unlike the
covariant Green-Schwarz formalism, this new formalism is easily quantized
with a BRST operator and tree-level scattering amplitudes
have been evaluated in a manifestly covariant manner.

In this paper, the cohomology of the BRST operator in the pure spinor
formalism is shown
to give the usual light-cone Green-Schwarz spectrum. Although the
BRST operator does not directly involve the Virasoro constraint, this
constraint emerges after expressing the pure spinor variable in terms
of SO(8) variables.

\Date {June 2000}

\newsec{Introduction}

Ever since the light-cone Green-Schwarz (GS) superstring formalism was
constructed \ref\lc{M.B. Green and J.H. Schwarz, 
{\it Supersymmetrical Dual String Theory}, Nucl. Phys. B181 (1981) 502.}, 
physicists have searched for a manifestly covariant
version of the formalism. Such a formalism
would have the advantage over the Ramond-Neveu-Schwarz formalism that
scattering amplitudes could be computed in a manifestly super-Poincar\'e
covariant manner.
Although there exists a classical covariant GS description
of the superstring \ref\GS{M.B. Green and J.H. Schwarz,
{\it Covariant Description of Superstrings}, Phys. Lett. B136 (1984) 367.}.
quantization problems have prevented this description
from being used to compute scattering amplitudes. Recently, a new
super-Poincar\'e covariant formalism for the superstring was constructed
using pure spinor worldsheet variables in addition to the usual
GS variables. Unlike all other covariant versions of the GS superstring,
this pure spinor formalism is easy to quantize and was used to compute
spacetime-supersymmetric tree
amplitudes involving an arbitrary number of external massless states
\ref\one{N. Berkovits, {\it Super-Poincar\'e Covariant Quantization of the
Superstring}, JHEP 04 (2000) 018, hep-th/0001035.}\ref\bc{N. Berkovits
and B.C. Vallilo, {\it Consistency of Super-Poincar\'e Covariant
Superstring Tree Amplitudes}, hep-th/0004171.}.

Physical states in this new formalism are defined as states in the cohomology
of the nilpotent operator 
$$Q=\int d\s \l^\a(z) d_\a(z)$$
where $\l^\a$ is the pure spinor variable and $d_\a$ is the worldsheet
variable for the spacetime-supersymmetric derivative \ref\csm
{W. Siegel, {\it Classical Superstring Mechanics}, Nucl. Phys. B263 (1986)
93.}. Although it is easy to check that the massless states in the
cohomology of $Q$ are those of ten-dimensional super-Yang-Mills \one
\ref\howe{P.S. Howe, {\it Pure Spinor Lines in Superspace and
Ten-Dimensional Supersymmetric Theories}, Phys. Lett. B258 (1991) 141.},
it is a bit mysterious how the correct massive
spectrum can be obtained since $Q$ does not directly involve the 
Virasoro constraint.
In this paper, this mystery will be resolved and it will be
shown that the cohomology of $Q$ indeed reproduces the desired light-cone
GS spectrum.

As will be discussed in section 2, the first step in resolving the mystery
is to express the pure spinor variable $\l^\a$ in terms of SO(8)
variables. The pure spinor constraint $\l\g^\mu\l=0$ implies that
$(\g^+\l)^a = s^a$ and 
$(\g^-\l)^\ad = \s_j^{a\ad} s^a v^j$ where $s^a$ is a null SO(8)
spinor satisfying $s^a s^a=0$ and $v^j$
is an unconstrained SO(8) vector.
In terms of $s^a$ and $v^j$,
$$Q= \int d\s ~ s^a [(\g^- d)^a + \s_j^{a\ad} v^j (\g^+ d)^\ad]$$
plus contributions from
an infinite chain of ghosts-for-ghosts
coming from the gauge invariance 
$\d v^j = \s^j_{a\ad} s^a \e^{\ad}$.

The second step in resolving the mystery is to enforce the first-class
constraint $s^a s^a=0$ by modifying the BRST operator to
$$Q'=Q +\int d\s [-b s^a s^a + c (\half\Pi^- + v^j \Pi^j +\half
v^j v^j \Pi^+)]$$
where $(b,c)$ is the ghost and anti-ghost for the $s^a s^a$ constraint, 
$\Pi_\mu=
\p x_\mu -\half \t\g_\mu\p\t$ is the spacetime-supersymmetric
momentum, and the term
$ c (\half\Pi^- + v^j \Pi^j + \half 
v^j v^j \Pi^+)$ is required for nilpotency of $Q'$.
As will be argued in section 3, $Q'$ has the same cohomology as $Q$ and is 
SO(9,1) super-Poincar\'e invariant.

Finally, it will be shown in section 4 that the cohomology of $Q'$  
reproduces the desired light-cone GS spectrum. 
Note that if one shifts $v^j\to v^j - \Pi^j/\Pi^+$ in $Q'$, 
$$Q'=\int d\s [(\Pi^+)^{-1} s^a (\Pi_\mu(\g^+\g^\mu d)^a + ...) -
b s^a s^a + (\Pi^+)^{-1}
c (-\half\Pi^\mu \Pi_\mu + ...)].$$
One can recognize 
$\Pi_\mu \g^+\g^\mu d$ as the first-class part of the GS fermionic
constraints and
$-\half\Pi^\mu\Pi_\mu$ as the GS Virasoro constraint.
The dependence of $Q'$ on $v^j$ and the
infinite chain of ghosts-for-ghosts is responsible for imposing the
second-class part of the GS fermionic constraints.
This use of an infinite set of fields for imposing second-class constraints
resembles the treatment of chiral bosons in
\ref\mc{B. McClain, Y.S. Wu and F. Yu,
{\it Covariant Quantization of Chiral Bosons and OSp(1,1/2) Symmetry},
Nucl. Phys. B343 (1990) 689\semi C. Wotzasek, {\it The Wess-Zumino Term
for Chiral Bosons}, Phys. Rev. Lett. 66 (1991) 129.} and self-dual 
four-forms in \ref\sft{N. Berkovits,
{\it Manifest Electromagnetic Duality in Closed
Superstring Field Theory}, Phys. Lett. B388 (1996) 743, hep-th/9607070.}.

\newsec{Construction of $Q$ using SO(8) Variables}

\subsec{Review of massless cohomology}

Physical states in the pure spinor formalism of the superstring are
defined as ghost-number one states in the cohomology of 
\eqn\Qdef{Q=\int d\s \l^\a(z) d_\a(z)}
where 
\eqn\dd{d_\a = p_\a +\half
\g^\mu_{\a\b}\p x_\mu \t^\b +{1\over 8}\g^\mu_{\a\b}\g_{\mu\,\g\d}
\t^\b\t^\g\p\t^\d}
is the worldsheet variable for the supersymmetric derivative \csm,
$p_\a$ is the conjugate momentum to $\t^\a$, and $\l^\a$ is a worldsheet
variable carrying $+1$ ghost number and satisfying the pure spinor
constraint 
\eqn\pure{\l^\a(z)\g^\mu_{\a\b}\l^\b(z)=0} 
for $\mu=0$ to 9.
Since $d_\a(y)d_\b(z)\to  (y-z)^{-1} \g^\mu_{\a\b} \Pi_\mu(z)$ where
$\Pi^\mu = \p x^\mu - \half\t^\a \g^\mu_{\a\b} \p\t^\b$,
$Q$ is nilpotent.

To see that the open superstring\foot{Although only the open superstring
will be discussed in this paper, all results are easily generalized
to the heterotic and closed superstrings.}
massless states are correctly reproduced
by the cohomology of the zero modes of $Q$, recall that on-shell 
super-Yang-Mills can be described by a spinor superfield $A_\a(x,\t)$
satisfying $D_\a (\g^{\mu_1 ... \mu_5})^{\a\b}
 A_\b =0$ for any five-form direction
$\mu_1 ... \mu_5$ where $D_\a = {\p\over{\p\t^\a}}-\half
\t^\b\g^\mu_{\a\b}\p_\mu$
\ref\beng{
W. Siegel, {\it Superfields in Higher-Dimensional Spacetime},
Phys. Lett. 80B (1979) 220\semi 
B.E.W. Nilsson, {\it Off-shell Fields for the Ten-Dimensional
Supersymmetric Yang-Mills Theory}, Gotenburg preprint 81-6 (Feb. 1981),
unpublished\semi B.E.W. Nilsson, {\it Pure Spinors as Auxiliary Fields
in the Ten-Dimensional Supersymmetric Yang-Mills Theory}, Class. Quant. Grav.
3 (1986) L41.}
\ref\wit{E. Witten, {\it Twistor-like Transform in Ten Dimensions},
Nucl. Phys. B266 (1986) 245.}
\howe.
Using the gauge invariance $\d A_\a = D_\a \Omega$, $A_\a$ can be gauge-fixed
to 
\eqn\gauge{A_\a(x,\t) =  a_\mu(x) \g^\mu_{\a\b}\t^\b + 
\xi^\g(x) \g^\mu_{\a\b}\g_{\mu\,\g\d}\t^\b\t^\d + ...}
where $a_\mu(x)$ and $\xi^\a(x)$ are
the linearized on-shell gluon and gluino of
super-Yang-Mills and the component fields in $...$ are auxiliary fields
which can be expressed in terms of $a_\mu$ and $\xi^\g$.

Since a massless
vertex operator only depends on the worldsheet zero modes, $V=\l^\a A_\a(x,\t)$
for some $A_\a(x,\t)$. But $QV=0$ implies that $\l^\a\l^\b D_\a A_\b=0$,
which can be decomposed into
$(\l\g^\mu\l) (D\g_\mu A)+
(\l\g^{\mu_1 ...\mu_5}\l) (D\g_{\mu_1 ... \mu_5} A)=0.$ Since
$\l\g^\mu\l=0$, $QV=0$ implies the desired equation that
$D\g_{\mu_1 ... \mu_5} A=0.$ Furthermore, the gauge invariance
$\d V=Q\Omega =\l^\a D_\a \Omega$ reproduces the desired gauge
transformation $\d A_\a = D_\a \Omega$.

So the cohomology of the zero modes of $Q$ correctly reproduces on-shell
super-Yang-Mills.
However, since $Q$ does not directly involve the Virasoro constraint,
it is a bit mysterious how the mass-shell condition for the physical
massive states is implied by $QV=0$. As mentioned in the introduction,
the first step to resolving this mystery is to express the pure spinor
$\l^\a$ in terms of SO(8) representations.

\subsec{ SO(8) parameterization of a pure spinor}

An SO(9,1) spinor $\l^\a$ satisfying $\l\g^\mu\l=0$ 
contains eleven independent complex degrees of freedom.
Together with their conjugate momenta, these eleven degrees of freedom
contribute $+22$ to the central charge which cancels the sum of the
central charge contributions of $+10$ from $x^\mu$ and 
$-32$ from $(\t^\a,p_\a)$. 

A convenient parameterization of $\l^\a$
is\one 
\eqn\lam{\lambda^+ = \g
, \quad \lambda_{AB} = \g u_{AB},\quad
\lambda^A =  -{1\over 8}\g\e^{ABCDE} u_{BC} u_{DE}}
where $A=1$ to 5, $u_{AB}=-u_{BA}$ parameterizes the ten-dimensional
complex space $SO(10)/U(5)$, and $\l^\a$ has been decomposed
(after Wick rotation) into its U(5) components. However, since
$\g$ is an overall scale parameter, this parameterization is singular
when the $\l^+$ component of $\l$ vanishes. Since physical states
can exist with vanishing $\l^+$, the parameterization of \lam\
is inappropriate
for computations of cohomology.\foot{For example, the massless vertex 
operator $V=\l^\a A_\a$ has physical degrees of freedom when $\l^+=0$.
For this reason, the fact that $V=\{Q, \g^{-1}\t^+ V\}$
does not imply that $V$ is BRST-trivial since $\g^{-1} \t^+ V$ is not a 
well-defined operator.}

An alternative parameterization of $\l^\a$ is in terms of its SO(8)
components $(\g^+\l)^a$ and $(\g^-\l)^\ad$ where $\g^\pm=\half
(\g^0\pm\g^9)$
and $(a,\ad)=1$ to 8 are chiral and anti-chiral SO(8) spinor indices.
The constraint 
$\l\g^-\l=0$ implies that $s^a=(\g^+\l)^a$ satisfies $s^a s^a=0$. 
Furthermore, the constraint
$\l\g^j\l=0$ implies that $(\g^-\l)^\ad=\s_j^{a\ad} v^j s^a$ 
for some SO(8) vector $v^j$ where $\s_j^{a\ad}$ are the SO(8) Pauli
matrices satisfying $\s_{(j}^{a\ad}\s_{k)}^{b\ad}=2\d_{jk}\d^{ab}$. 
One can check that the constraint
$\l\g^+\l=0$ implies no further conditions on $s^a$ and $v^j$.
So the eleven degrees of freedome of $\l^\a$ can be parameterized by
the seven degrees of freedom of a null spinor $s^a$ together with
the eight degrees of freedom of $v^j$ as
\eqn\la{(\g^+\lambda)^a = s^a , \quad 
(\g^-\lambda)^\ad = \s_j^{a\ad} v^j s^a.}
Unlike the U(5) parameterization
of $\lam$, this SO(8) parameterization is singular only when all eight
components of $(\g^+\l)^a$ are zero. However, there are no physical states
with vanishing 
$(\g^+\l)^a$, 
so the parameterization of \la\ 
is appropriate for computing the
cohomology.\foot{For example, the gauge invariance $\d A_\a = D_\a \Omega$
implies that one can choose the gauge $(\g^+ A)^\ad$ for the super-Yang-Mills
spinor prepotential \wit. In this gauge, the massless vertex operator
$V=\l^\a A_\a$ vanishes when $(\g^+\l)^a=0$. One expects that a 
similar gauge choice is possible for physical massive vertex operators
such that they vanish when $(\g^+\l)^a=0$.}

Since \la\ is invariant under 
\eqn\gt{\d v^j=\s^j_{a\ad}s^a \e^{\ad}}
for arbitrary $\e^\ad$, this parameterization of $\l^\a$
has a gauge invariance
which needs to be correctly treated. This can be done in the usual BRST
manner by introducing a fermionic ghost SO(8) spinor variable $t^\ad$.
However, since 
$\d \e^\ad=\s^j_{a\ad}s^a y^j$ leaves the gauge transformation of \gt\
unchanged, one also needs to introduce a bosonic ghost-for-ghost
SO(8)
vector variable
$v^j_{(1)}$. This line of reasoning continues ad infinitum to produce
an infinite chain of bosonic SO(8) vectors, $v^j_{(0)}, v^j_{(1)}, ...$,
and 
an infinite chain of fermionic SO(8) spinors, $t^\ad_{(0)}, t^\ad_{(1)}, ...$,
where the original $v^j$ and $t^\ad$ variables have been relabeled as
$v^j_{(0)}$ and $t^\ad_{(0)}$.

Since $s^a, v^j_{(n)}$ and $t^\ad_{(n)}$ carry zero conformal weight,
they contribute (together with their conjugate momenta)
$+2(7+8-8+8-8+ ...)$ to the central charge.
Using the regularization familiar from $\kappa$-symmetry computations
\ref\ghosts{
W. Siegel, {\it Lorentz Covariant Gauges for Green-Schwarz 
Superstrings}, talk at Strings '89, College Station Workshop (1989) 211\semi
S.J. Gates Jr., M.T. Grisaru, U. Lindstrom,  M. Rocek, W. Siegel and
P. van Nieuwenhuizen, 
{\it Lorentz Covariant Quantization of the Heterotic Superstring},
Phys. Lett. B225 (1989) 44\semi
R.E. Kallosh, {\it Covariant Quantization of Type IIA,B
Green-Schwarz Superstring}, Phys. Lett. B225 (1989) 49\semi
M.B. Green and C.M. Hull, {\it Covariant Quantum Mechanics of the
Superstring}, Phys. Lett. B225 (1989) 57.} that
\eqn\regulone{8-8+8-8+ ... = \lim_{x\to 1} 8 (1-x^2+x^3-x^4 + ...) =
 \lim_{x\to 1} 8 (1+x)^{-1} = 4,}
one recovers the desired $+22$ contribution to the central charge.

Including the contribution of the ghost-for-ghosts, the 
BRST charge is 
$Q=\int d\s s^a G^a$ where 
\eqn\defG{G^a = (\g^- d)^a + \s_j^{a\ad} [v^j_{(0)} (\g^+ d)^\ad
+\sum_{n=0}^\infty (w^j_{(n)} t^\ad_{(n)} + v^j_{(n+1)} u^\ad_{(n)})],}
$w^j_{(n)}$ is the conjugate momentum to $v^j_{(n)}$, and
$u^\ad_{(n)}$ is the conjugate momentum to $t^\ad_{(n)}$.
Note that $Q^2=0$ since $s^a s^a=0$ and $G^a(y) G^b(z)\to 2\d^{ab} (y-z)^{-1}
T(z)$ where
\eqn\defT{T=\half\Pi^- + v^j\Pi^j +\half
v^j v^j \Pi^+ + t^\ad_{(0)} (\g^+ d)^\ad
+\sum_{n=0}^\infty (v^j_{(n+1)} w^j_{(n)} + t^\ad_{(n+1)} u^\ad_{(n)})}
and $\Pi^\pm = \Pi^0 \pm \Pi^9$.

Although $G^a(y) G^b(z)\to 2\d^{ab} (y-z)^{-1} T(z)$ suggests an $N=8$
super-Virasoro algebra, $G^a$ and $T$ are {\it not} super-Virasoro
generators since, for example, $G^a$ and $T$ have $+1$ conformal
weight and $T$ has no singular OPE's with either
$G^a$ or $T$. Nevertheless, the resemblance with an $N=8$ algebra
suggests that the BRST operator $Q$ can be modified to 
\eqn\brs{Q'=\int d\s (s^a G^a + c T - b s^a s^a)} 
where $(b,c)$ are fermionic ghosts of conformal weight $(1,0)$.
It will be shown in the following section that $Q'$ indeed
has the
same cohomology as $Q$.

\newsec{BRST treatment of the $s^a s^a=0$ Constraint}

\subsec{ Equivalence of cohomology of $Q$ and $Q'$}

Since the constraint $s^a s^a$ is included in the 
BRST operator $Q'$ of \brs, one expects that all eight components of $s^a$
can be treated as independent degrees of freedom in the `off-shell'
Hilbert space of $Q'$. Note that
this does not affect the central charge computation
since the $-2$ contribution of the $(b,c)$ ghosts cancels the $+2$
contribution of the extra degree of freedom in $s^a$ and its conjugate
momentum.

It will now be argued that the cohomology of $Q'$ with $s^a$ unconstrained
is equivalent to the cohomology of $Q$ with $s^a$ constrained to satisfy
$s^a s^a=0$.
Consider a state $V$ annihilated by
$Q$ up to terms involving $s^a s^a$, i.e. $Q V= s^a s^a W$
for some $W$. Then $Q^2 = s^a s^a T$ implies that $QW=TV$.
Using this information, one can check that 
the operator $V'=V+c W$ is annihilated by $Q'$.
Furthermore, if $V$ is BRST-trivial up to terms involving $s^a s^a$, 
i.e. $V= Q\Omega  + s^a s^a Y$ for some
$Y$, then $V+cW =Q'(\Omega-cY)$ so $V'$ is also BRST-trivial.

At the end of section 4, it will be shown that all
physical states (with non-zero $P^+$) in the cohomology of $Q'$
can be written in the form $V'=V+cW$ for some $V$ and $W$. Reversing
the arguments of the previous paragraph, one learns that $V$ is
in the cohomology of $Q$ up to terms involving $s^a s^a$. This proves
equivalence of the cohomologies.

\subsec{Super-Poincar\'e invariance of $Q'$}

Although $Q'$ of \brs\ is expressed in terms of 
SO(8) variables, it will now be argued that $Q'$
is invariant under SO(9,1) transformations.
Since $Q'$ is manifestly spacetime-supersymmetric, this implies
the super-Poincar\'e invariance of $Q'$.
In terms of SO(8) representations, the pure spinor contribution
to the SO(9,1) Lorentz currents is
\eqn\lore{N^{jk}=\half s^a (\s^{jk})_{ab} r^b +\sum_{n=0}^\infty
[v_{(n)}^{[j} w^{k]}_{(n)} +\half 
t_{(n)}^\ad (\s^{jk})_{\ad\bd} u^\bd_{(n)}],}
$$N^{j+}= w^j_{(0)},$$
$$N^{+-}=bc -\half s^a  r^a +\sum_{n=0}^\infty [ (n+1)
v_{(n)}^j w^j_{(n)} + (n+{3\over 2})
t_{(n)}^\ad w^\ad_{(n)}], $$
$$N^{j-}=-3\p v_{(0)}^j - v^k_{(0)} N^{jk}-
v^j_{(0)} N^{+-} -\half  v^k_{(0)} v^k_{(0)} w^j_{(0)}
+ v^j_{(0)} v^k_{(0)} w^k_{(0)} +
\half c \s^j_{a\ad} t_{(0)}^\ad r^a + F^{j -},$$
where $r^a$ is the conjugate momentum to $s^a$ and 
it should be possible to determine
the term $F^{j -}$ 
by requiring that 
$$[\int d\s N^{j -}~ , ~
\sum_{n=0}^\infty  \big( s^a \s_j^{a\ad}
(w^j_{(n)} t^\ad_{(n)} + v^j_{(n+1)} u^\ad_{(n)})
+c (v^j_{(n+1)} w^j_{(n)} + t^\ad_{(n+1)} u^\ad_{(n)})\big) - s^a s^a b ~]=0.$$
Note that $[s^a, ~\s_j^{a\ad} v^j_{(0)} s^a + c t_{(0)}^\ad]$ transform as the
sixteen components of an SO(9,1) spinor and $[-\half(c
+ c v^k_{(0)} v^k_{(0)}),~cv^j_{(0)}, ~
-\half (c-
c v^k_{(0)} v^k_{(0)})]$ transform as the ten components of an SO(9,1) vector,
so the terms 
$[s^a(\g^- d)^a + (\s_j^{a\ad} s^a v^j_{(0)}+ct_{(0)}^\ad) (\g^+ d)^\ad]$ and
$[\half c\Pi^- + c v^j_{(0)} \Pi^j +
\half c v^k_{(0)} v^k_{(0)} \Pi^+]$ in $Q'$ are easily seen to be Lorentz 
invariant.

Furthermore, one can check (up to the determination of $F^{j-}$)
that $N^{\mu\nu}$ of \lore\ satisfies the OPE
\eqn\nope{N^{\mu\nu}(y) N^{\rho\sigma}(z) \to
{{\eta^{\rho[\nu} N^{\mu]\s}(z) -
\eta^{\s[\nu} N^{\mu]\rho}(z) }\over {y-z}} - 3
{{\eta^{\mu\s} \eta^{\nu\rho} -
\eta^{\mu\rho} \eta^{\nu\s}}\over{(y-z)^2}}  }
where the factor of 3 in the double pole comes from the pure spinor
condition and is crucial for equivalence with the Lorentz generators
in the Ramond-Neveu-Schwarz formalism for the superstring \one.
For example, the double pole in $N^{jk}$ with $N^{jk}$ gets a contribution
of $+2$ from the first term in $N^{jk}$ and a contribution of 
$+2-2+2-2+...$ from the remaining terms.
Using the regularization of \ghosts, 
\eqn\regtwo{+2-2+2-2+... = \lim_{x\to 1} 2 (1+x)^{-1} =1,}
so the total double pole contribution is $+3$ as desired.
Simlilarly, the double pole of $N^{+-}$ with $N^{+-}$ gets
a contribution of $+1$ from the first term, $-2$ from the second term,
and $-2(2^2-3^2 +4^2 -5^2 + ...)$ from the remaining terms.
This last expression can be regularized using the formula
\eqn\formu{\sum_{n=0}^\infty n^2 (-x)^n = 2 (1+x)^{-3} -3 (1+x)^{-2}
+ (1+x)^{-1},}
which can be obtained by taking derivatives of the formula
$\sum_{n=0}^\infty (-x)^n = (1+x)^{-1}$. So
\eqn\regulth{ 2^2-3^2 +4^2 -5^2 + ... = 1 +
\lim_{x\to 1}\sum_{n=0}^\infty n^2 (-x)^n = 1 + \lim_{x\to 1}
[2 (1+x)^{-3} -3 (1+x)^{-2}
+ (1+x)^{-1}] = 1,}
implying that the sum of the double pole contributions is $-3$ as desired.

So $Q'$ has been shown to be a super-Poincar\'e
invariant operator whose
cohomology is equivalent to that of $Q=\int d\s \l^\a d_\a$. The cohomology
of $Q'$ will now be computed to be the light-cone GS spectrum.

\newsec{Evaluation of Cohomology of $Q'$}

\subsec{ Light-cone operators}

As mentioned earlier, $Q'$ resembles the BRST operator for an $N=8$
super-Virasoro algebra. This can be made more evident by shifting
$v^j_{(0)}\to v^j_{(0)} - (\Pi^+)^{-1} \Pi^j$ (where the zero mode
of $\Pi^+$ is assumed to be non-vanishing), so that
$G^a= (\Pi^+)^{-1} \Pi_\mu (\g^+\g^\mu d)^a + ...$ and
$T= -\half (\Pi^+)^{-1} \Pi_\mu \Pi^\mu + ...$. The first term in $G^a$
can be recognized as the first-class part of the fermionic GS constraint
and the first term in $T$ can be recognized as the GS Virasoro constraint.
As will be explained below, the second-class part of the fermionic GS
constraint will be implied by the infinite ghost-for-ghost dependence of $Q'$
in a manner similar to the treatment of chiral bosons in \mc\ and
self-dual four-forms in \sft.

To compute the cohomology of $Q'$, it is useful to first write $Q'=Q_1 +Q_2$
where 
\eqn\brstwo{Q_1=\int d\s [ s^a (\g^- p)^a + 
\half c (\p x^- - P^-) ],\quad Q_2=
Q'-Q_1,}
and $P^\mu$ is the zero mode of $\p x^\mu$. If $[(\g^- p)^a,(\g^+\t)^a]$
are assigned charge $(+1,-1)$, 
$[(\p x^- -P^-), (\p x^- - P^+)] $ are 
assigned charge $(+1,-1)$, and all other variables are assigned zero charge,
then $Q_1$ has charge $+1$ and all terms in $Q_2$ have non-positive charge.

So the cohomology of $Q'$ is given by the cohomology of $Q_2$ restricted
to operators in the cohomology of $Q_1$ \ref\priv{E. Witten,
private communication.}. But the 
cohomology of $Q_1$ consists of operators which are independent of 
$[(\g^- p)^a,(\g^+\t)^a,
(\p x^- -P^-), (\p x^+ - P^+), s^a, r^a ] $  and the non-zero modes of
$(b,c)$. So the only term in $Q_2$ which survives in the cohomology of
$Q_1$ is $c_0 (T_0 +\half P^-)$ where
\eqn\deftz{T_0 = 
\int d\s [ -\half\t\g^-\p\t + v^j_{(0)} \p x^j
+ \half v^j_{(0)}v^j_{(0)} P^+ }
$$ + t_{(0)}^\ad (\g^+ p +\half P^+ \g^-\t)^\ad +
\sum_{n=0}^\infty (v^j_{(n+1)} w^j_{(n)} +t^\ad_{(n+1)} u^\ad_{(n)})].$$

A general operator in the cohomology of $Q_1$ can be written as
${\cal O} = A + b_0 B + c_0 C$ where
$A,B,C$ are independent of 
$[(\g^- p)^a,(\g^+\t)^a,
(\p x^- -P^-), (\p x^+ - P^+), s^a, r^a ] $.
Furthermore, $[Q_2,{\cal O}]=0$ implies that $[T_0+\half P^-,A]=0$ and
$B=0$. Finally, the gauge invariance $\d{\cal O}=Q_2\Omega$
implies $\d C=[T_0 +\half P^-,\Omega]$, so the cohomology associated with $C$
is related to that of $A$
by the usual doubling phenomenon associated with the $c_0$ ghost.
So the cohomology of $Q'$ can be evaluated by solving
the equation $[T_0+\half P^-,A]=0$.

Note that this same result can be obtained by using `old covariant
quantization' where one ignores the $(s^a,r^a)$ and $(c,b)$ ghosts.
Using this method, one first uses $G^a$ and the non-zero modes of $T$ to
gauge away $(\g^+\t)^a$ and $\p x^+$. Requiring that the
operator $A$ commutes with $G^a$ and $T$ fixes $(\g^- p)^a$ and $\p x^-$
in terms of the remaining light-cone variables and implies that 
$[T_0+\half P^-,A]=0$.

Since $T_0$ is quadratic in the remaining worldsheet variables,
any operator $A$ satisfying $[T_0+\half P^-,A]= 0$ can be constructed from
products of linear combinations of the variables, $a_N$, which satisfy
$[T_0, a_N]= N a_N$ for some $N$. Then $[T_0 +\half P^-,A]=0$
implies the mass-shell condition
that $-\half P^-$ is equal the sum of the eigenvalues in the
product. For convenience, a Lorentz frame will be chosen where $P^+$
is a non-zero fixed constant and $P^j=0$
for $j=1$ to 8.

One can easily check that $[T_0, y^j] = - (P^+)^{-1}\p y^j$ and
$[T_0, q^\ad] = - (P^+)^{-1}\p q^\ad$ where
\eqn\eigen{y^j = \p x^j +  \sum_{n=0}^\infty (P^+)^{-n-1}
\p^{n+1} w_{(n)}^j,\quad
q^\ad = (\g^+ p -\half P^+ \g^-\t)^\ad + 
\sum_{n=0}^\infty (P^+)^{-n-1}\p^{n+1} u_{(n)}^\ad.}
So the $M^{th}$ mode of $y^j$ and $q^\ad$ are eigenvectors
of $T_0$ which
carry eigenvalue $N=-M/P^+$.
In fact, as will now be shown, these are the only normalizable
eigenvectors of $T_0$ which can be constructed from linear combinations
of the remaining variables. 

First, suppose one has a bosonic eigenvector of $T_0$ of the form
\eqn\boson{a_N =
\int d\s[ f_N^j \p x^j +\sum_{n=0}^\infty (g_{N(n)}^j v_{(n)}^j +
h_{N(n)}^j w_{(n)}^j )]}
where $(f_N^j, g_{N(n)}^j, h_{N(n)}^j)$ are coefficients of the eigenvector.
Then $[T_0,a_N]=N a_N$ implies that
\eqn\bosf{ -h_{N(0)}^j=N f_N^j,
\quad -P^+ h_{N(0)}^j - \p f_N^j =N g_{N(0)}^j, }
$$
g_{N(n)}^j= N g_{N(n+1)}^j,\quad 
-h_{N(n+1)}^j= N h_{N(n)}^j.$$
Using the normalizability condition that 
$\int d\s[ f_N^j f_N^j +\sum_{n=0}^\infty g_{N(n)}^j h_{N(n)}^j ]$
is finite, one finds that the only normalizable solution of \bosf\ is
$$\p f_N^j = P^+ N f_N^j, \quad g_{N(n)}^j=0,\quad h_{N(n)}^j= (-N)^{n+1}
f_N^j,$$
which is the $(-P^+ N)^{th}$ mode of the eigenvector $y^j$ of \eigen.

Second, suppose one has a fermionic eigenvector of $T_0$ of the form
\eqn\fermion{a_N =
\int d\s[ j_N^\ad (\g^+ p)^\ad + k_N^\ad (\g^-\t)^\ad 
+\sum_{n=0}^\infty (l_{N(n)}^\ad t_{(n)}^\ad +
m_{N(n)}^\ad u_{(n)}^\ad )]}
where $(j_N^\ad,k_N^\ad, l_{N(n)}^\ad, m_{N(n)}^\ad)$ 
are coefficients of the eigenvector.
Then $[T_0,a_N]=N a_N$ implies that
\eqn\fermf{ -m_{N(0)}^\ad=N j_N^\ad, \quad
-\half P^+ m_{N(0)}^\ad - \p j_N^\ad =N k_N^\ad,\quad
\half P^+ j_N^\ad +  k_N^\ad = N l_{N(0)}^\ad,}
$$
l_{N(n)}^\ad= N l_{N(n+1)}^\ad,\quad 
-m_{N(n+1)}^\ad= N m_{N(n)}^\ad.$$
Using the normalizability condition that 
$\int d\s[ j_N^\ad k_N^\ad +\sum_{n=0}^\infty l_{N(n)}^\ad m_{N(n)}^\ad ]$
is finite, one finds that the only normalizable solution of \fermf\ is
$$\p j_N^\ad = P^+ N j_N^\ad, \quad
k_N^\ad=-\half P^+ j_N^\ad,\quad l_{N(n)}^\ad=0,\quad
m_{N(n)}^\ad= (-N)^{n+1} j_N^\ad,$$
which is the $(-P^+ N)^{th}$ mode of the eigenvector $q^\ad$ of \eigen.

So any operator satisfying $[T_0 +\half P^-,A]=0$ 
can be expressed as a product of
the modes of $y^j$ and $q^\ad$ multiplied by $e^{iP^- x^+}$ where
$\half P^+ P^-$ 
is the sum of the mode numbers. By acting on a `ground state' with
non-zero $P^+$, these light-cone operators will now be used to
construct physical states in the cohomology of $Q'$.

\subsec{Physical states}

Using the usual DDF construction \ref\ddf{E. DelGiudice, P. DiVecchia
and S. Fubini, {\it General Properties of the Dual Resonance Model},
Ann. Phys. 70 (1972) 378.}, the light-cone operators $y^j$ and $q^\ad$
of \eigen\ can be extended to operators $\hat y^j$ and $\hat q^\ad$
which commute with $G^a$ and $T$, and therefore commute with $Q'$.
Although $\hat y^j$ and $\hat q^\ad$ will depend on the variables
$\p x^+$ and $(\g^+\t)^a$, they will be independent of the 
$(c,b)$ and $(s^a, r^a)$ ghosts. Any operator (with $P^+$ non-zero
and $P^j=0$) in the cohomology of $Q'$
can be constructed from products of modes of $\hat y^j$ and $\hat q^\ad$
multiplied by the appropriate factor of $e^{iP^- x^+}$.

Physical states in the cohomology of $Q'$ are constructed by acting
with these operators on a `ground state' with non-zero $P^+$ and $P^j=0$. 
Using the construction of section 3.1 together with
the massless vertex operator
of section 2.1, a suitable such ground state is
\eqn\suit{V'_0 = \big( s^a [(\g^- A)^a + \s_j^{a\ad} v_{(0)}^j (\g^+ A)^\ad]
+ c [ (D\g^- A) + v_{(0)}^j (D\g^j A) + v_{(0)}^j v_{(0)}^j (D\g^+ A)] \big)
e^{i P^+ x^-} }
where $A_\a$ is the on-shell super-Yang-Mills prepotential and $D_\a$
is the supersymmetric derivative.
This state is annihilated by all negative modes of $\hat y^j$ and
$\hat q^\ad$, and the zero mode of $\hat q^\ad$ acts as a spacetime
supersymmetry transformation on $V'_0$. 

So the physical states in the
cohomology of $Q'$ (with non-zero $P^+$ and $P^j=0$) can be represented by
\eqn\physf{V' = \prod_{j=1}^8 \prod_{\ad=1}^8 \prod_{m,n=1}^\infty
(\p^m\hat y^j)^{\a_m^j}
(\p^n\hat q^\ad)^{\b_n^\ad} e^{i P^- x^+} V'_0}
where $\half P^+
P^- = \sum_{j,\ad,m,n} ( m \a_m^j + n \b_n^\ad )$. This is the
usual light-cone GS spectrum. Note that all such states are of the
form $V' = V+cW$, which was needed in section 3.1 for proving equivalence
of the $Q$ and $Q'$ cohomologies.

\vskip 15pt

{\bf Acknowledgements:} I would like to especially thank Edward Witten
for suggesting the comparison with light-cone Green-Schwarz, for his
ideas concerning the cohomology computation, and for his collaboration
during the initial stage of this work. I would also like to thank
Cumrun Vafa for suggesting the importance of the $N=8$ algebra, 
Michael Bershadsky, Warren Siegel and Stefan Vandoren for useful
conversations, CNPq grant 300256/94-9
for partial financial support, and the univerisities of Caltech, Harvard
and SUNY at Stony Brook for their hospitality.
This research was partially conducted during the period the author was
employed by the Clay Mathematics Institute as a CMI Prize Fellow.

\listrefs

\end